\pdfoutput=1
\documentclass[aps,prl,showpacs,twocolumn]{revtex4}
\usepackage{graphicx}
\usepackage{amssymb}
\usepackage{amsmath}
\usepackage{color}
\usepackage{dsfont}
\usepackage{paralist}
\usepackage{bm}
\usepackage{verbatim}

\def\cL{{\cal L}}

\def\bk{{\bf k}}

\def\bp{{\bf p}}
\def\br{{\bf r}}

\def\hbsigma{\hat{\boldsymbol \sigma}}

\def\bnabla{{\boldsymbol \nabla}}

\def\holOne{\mathds{1}}
\def\holC{\mathds{C}}

\def\hbp{\hat{\bf p}}
\def\hv{\hat v}

\def\hsigma{\hat\sigma}

\def\be{\begin{equation}}
\def\ee{\end{equation}}
\def\bea{\begin{eqnarray}}
\def\eea{\end{eqnarray}}

\begin{document}

\title{Critical transport in weakly disordered semiconductors and semimetals}


\author{S.V.~Syzranov, L.~Radzihovsky, V.~Gurarie} 
\affiliation{Physics Department, University of Colorado, Boulder, CO 80309, USA}

\begin{abstract}
  {Motivated by Weyl semimetals and weakly doped semiconductors, we
    study transport in a weakly disordered semiconductor
    with a power-law quasiparticle dispersion $\xi_{\bf k}\propto
    k^\alpha$. We show, that in $2\alpha$ dimensions short-correlated
    disorder experiences logarithmic renormalisation from all energies
    in the band.  We study the case of a general dimension $d$ using a
    renormalisation group, controlled by an
    $\varepsilon=2\alpha-d$-expansion.  Above the critical dimensions,
    conduction exhibits a localisation-delocalisation phase
    transition or a sharp crossover (depending on the symmetries of
    the Hamiltonian) as a function of disorder strength. We utilise
    this analysis to compute the low-temperature conductivity in Weyl
    semimetals and weakly doped semiconductors near and below the
    critical disorder point.}
\end{abstract}



\date{\today}
\maketitle


Low-temperature conductivity in weakly disordered metals is usually dominated
by elastic scattering processes within a narrow shell of momentum states $\bk$
near the Fermi surface, $|E_\bk-E_F|\ll \tau^{-1}$, where $\tau$ is the elastic scattering time.
It is usually believed that scattering into the states outside of this shell is either negligible or may only renormalise the quasiparticle parameters near the Fermi surface, not leading to qualitatively new effects.

However, it is well known that Dirac-type quasiparticles in two dimensions (2D)
experience logarithmic renormalisation from elastic scattering into {\it all} states
corresponding to the linear spectrum,
as it has been shown long ago in the context of Ising models\cite{DotsenkoDotsenko},
degenerate semiconductors\cite{Fradkin1,Fradkin2}, the
integer Hall effect\cite{LudwigFisher}, d-wave superconductors\cite{Nersesyan:dwave},
{and topological insulators\cite{GoswamiChakravarti}}.
Recently, a similar renormalisation group (RG) description for transport in graphene
has been developed in Ref.~\onlinecite{AleinerEfetov}
and further discussed in Ref.~\onlinecite{OstrovskyGornyMirlin},
predicting a logarithmic dependence of physical observables on the electrostatically
tunable charge carrier concentration.



{In this paper we show that in a broad class of systems the transport of particles with kinetic energy $E$
experiences strong renormalisation from elastic scattering between all states in the band
provided the bandwidth is sufficiently large, $W\gg E$, which typically occurs in
semiconductors and semimetals.}

\begin{figure}[htbp]
	\centering
	\includegraphics[width=0.4\textwidth]{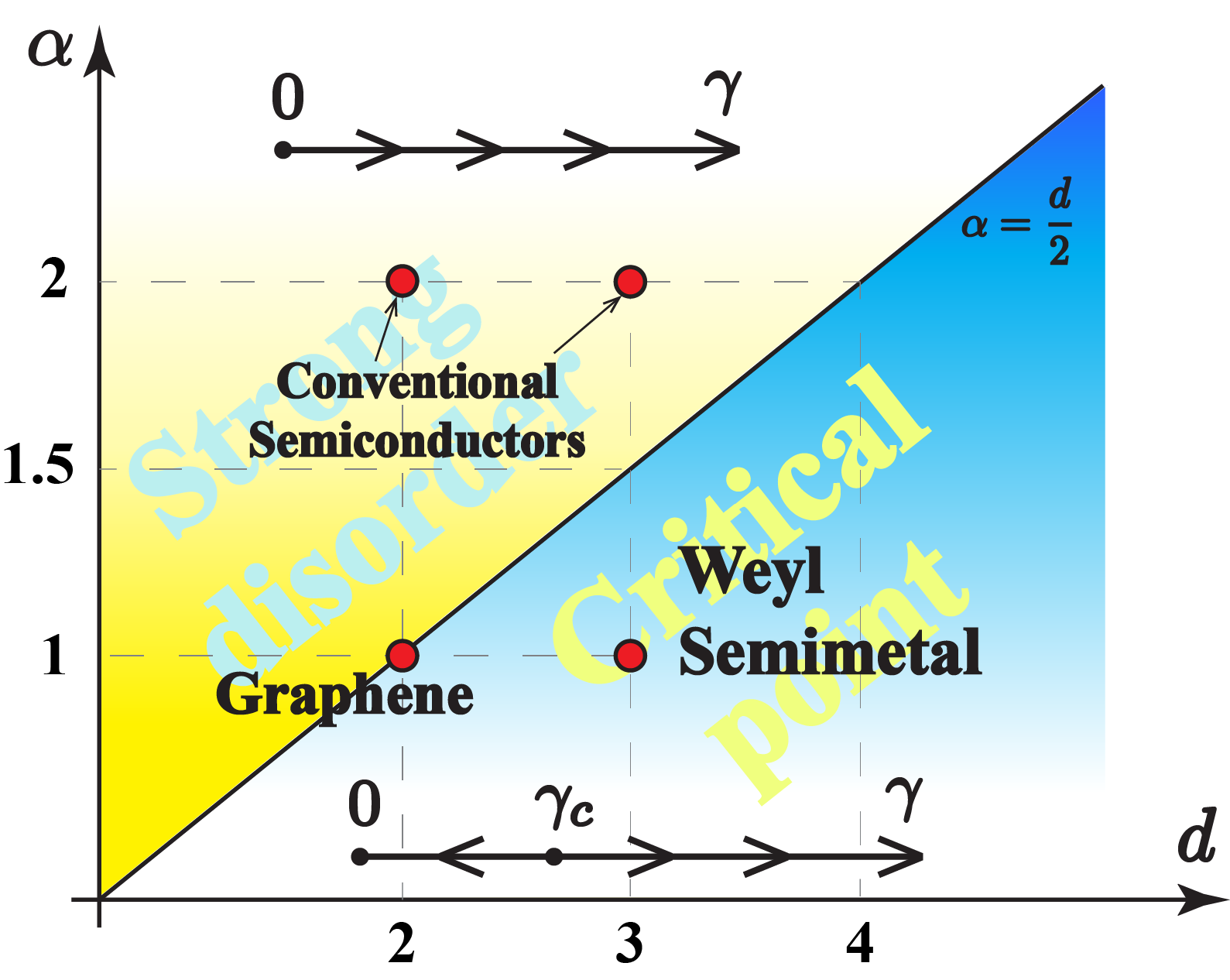}
	\caption{\label{PhaseDiagram}
	(Colour online) Critical
          behaviour of disorder in materials with a power-low
          quasiparticle dispersion $\xi_\bk\propto k^\alpha$ in $d$
          dimensions.  {\em Above} the $\alpha=d/2$ line the effects
          of disorder grow at low energies (the strong-disorder regime). 
          Materials
          {\em below} the line exhibit a critical point between the
          weak-disorder and strong-disorder regimes.
	}
\end{figure}

\begin{figure}[t!]
	\centering
	\includegraphics[width=0.4\textwidth]{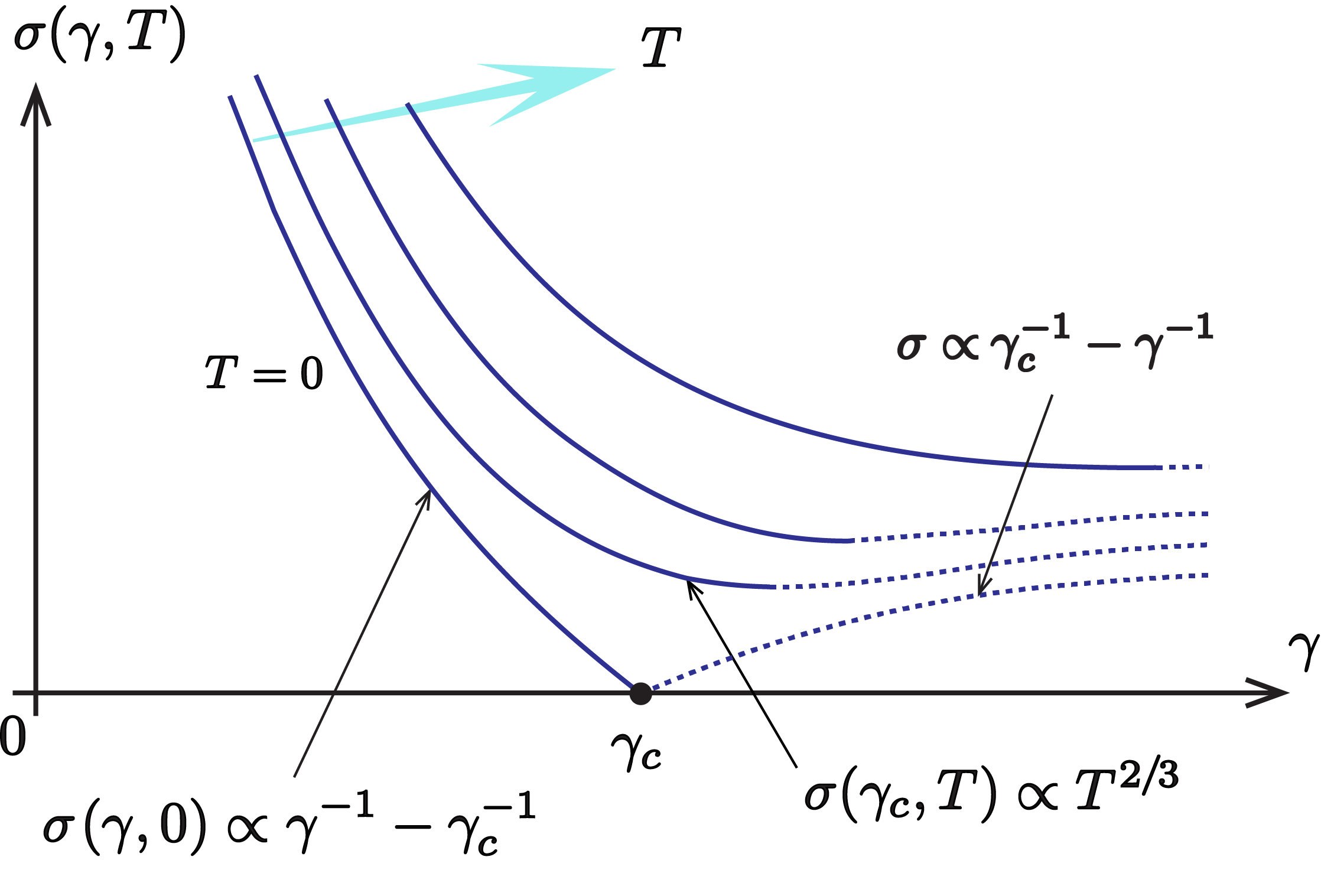}
	\caption{\label{WeylCondSurface}
	(Colour online) Conductivity of a Weyl semimetal at small finite doping $\mu$
	as a function of the disorder strength and temperature.
	The dashed parts of the $\sigma(\gamma,T)$ curves
	correspond to the strong-disorder regime and may be affected by
	the interference effects at large length scales	 not studied here[26]. 
	} 
\end{figure}

We study transport in a weakly disordered semiconductor or a semimetal with a power-law spectrum 
$\xi_{\bf k}\propto k^\alpha$ in a $d$-dimensional space. Our conclusions, regarding the
critical behaviour of a variety of systems, are summarised in Fig.~\ref{PhaseDiagram}.
In the critical dimension $d_c=2\alpha$, as exemplified by graphene\cite{AleinerEfetov,OstrovskyGornyMirlin} $(d=2, \alpha=1)$,
the disorder strength is subject to logarithmic renormalisations, qualitatively distinct from the weak-localisation corrections.
Transport in materials just below or above the critical dimension is accessible to a rigorous RG treatment,
supplemented
by an $\varepsilon$ expansion, where 
\begin{equation} \varepsilon =2\alpha-d.
\end{equation}
In the dimensions below critical, $d<d_c$,
the renormalised disorder strength increases at low energies. Above critical dimensions, the disorder
strength increases if its bare value exceeds a critical value, and flows to zero otherwise.
As a result, the conductivity $\sigma(\gamma)$ displays a
transition%
\footnote{\label{Fnote}
Depending on the symmetries of the Hamiltonian, this can be an Anderson localisation transition
or a sharp crossover. The details of the transition very close to the critical point
are determined by the interference 
effects on sufficiently large length scales [weak-(anti)localisation effects],
which have to be studied for particular materials by
other techniques, such as non-linear sigma-model\cite{Efetov:book}.
}
as a function
of the bare disorder strength, as summarised, for example, for a Weyl semimetal (WSM)
in Fig.~\ref{WeylCondSurface}.
Our conclusions persist even for quasiparticle Hamiltonians with non-trivial sublattice or valley
structures, as,
for example, in graphene
or a WSM.

{\it The model for semiconductors.}
Let us first consider critical behaviour in a $d$-dimensional semiconductor with the
band gap $2\Delta$, an isotropic spectrum 
$
	\xi_\bk=a k^\alpha
$	
in the conduction band,
and a trivial valley and sublattice structure. For simplicity, 
we consider a model in which the conductivity
is dominated by the electrons in the conduction band, e.g., due to a large
value of the gap $\Delta$ that exceeds the bandwidth or the scale $a r_0^{-\alpha}$,
with $r_0$ being the characteristic disorder correlation length, which then determines
the effective ultraviolet cutoff $K_0=r_0^{-1}$ of the theory\cite{Syzranov:UnconvLoc}.

We take the disorder potential $U(\br)$ to be weak, with zero-mean and
short-range correlated Gaussian statistics,
$
	\langle U(\br)U(\br^\prime)\rangle_{dis}=\gamma_0 K_0^\varepsilon\delta(\br-\br^\prime),
$	
characterised by the strength $\gamma_0$. The short-scale (ultraviolet
momentum) cutoff $K_0$ is set by the width of the conduction band.

{\it The dc conductivity} for zero temperature and the Fermi energy $E$
is given by the Kubo-Greenwood formula
\begin{align}
	\sigma_{ij}^{E}=\int d\br^\prime\:\:\mathrm{Tr}\left<\hv_{i\br} G^A(E,\br,\br^\prime) 
	\hv_{j\br^\prime} G^R(E,\br^\prime,\br)\right>_{dis},
	\label{Kubo}
\end{align}
where 
\( {\bf \hv}_{\br}=\alpha a(-i\bnabla_\br)^{\alpha-1} \)
is the velocity operator, $\hbar=e=1$, 
and the trace is taken with respect to (wrt) the discrete degrees of freedom
(spins, valleys, sublattices). All the energies $E$ are counted from the middle of the forbidden band,
where the chemical potential is located in an intrinsic semiconductor at $T=0$.
Conductivity at arbitrary temperature and doping level can be obtained from Eq.~(\ref{Kubo}) as
$	\sigma_{ij}=
	-\int dE \:n_F^\prime(E)\: \sigma_{ij}^{E}$, where $n_F(E)$ is the Fermi distribution
function.

The product of the advanced $G^A$ and retarded $G^R$ Green's functions in Eq.~(\ref{Kubo}),
averaged with respect to disorder, can be
written conveniently in the supersymmetric representation\cite{Efetov:book} as
\begin{align}
	&\langle\ldots\rangle_{dis}=
	\int {\cal D} {\overline\Psi}{\cal D}\Psi\ldots
	\exp{\left[-(\cL_0+\cL_{int})\right]}, \label{eq:gf}
	\\
	&\cL_0=i\int\overline{\Psi}
	\left[\lambda \left(E-\Delta\right)-\xi_{\hbp}-i\Lambda\cdot0\right]
	\Psi d\br,
	\\
	&\cL_{int}=\frac{1}{2}\gamma K^\varepsilon\int(\overline{\Psi}\Psi)^2d\br,
	\label{intL}
\end{align}
where $\Psi$ is a vector in $AR\otimes PH\otimes FB$ space; $AR$, $PH$, and $FB$ being, respectively, the
advanced-retarded, particle-hole, and fermion-boson subspaces;
$\Lambda=\hsigma_z^{AR}\otimes\holOne^{PH}\otimes\holOne^{FB}$, and 
$\overline{\Psi}=\Psi^\dagger\Lambda\equiv(\holC\Psi)^T$, where 
$\holC=\hsigma_z^{AR}\otimes(\hsigma_-^{PH}\otimes\holOne^{FB}-\hsigma_+^{PH}\otimes\hsigma_z^{FB})/2$ 
and $\hbp=-i\bnabla_\br$.  The
parameters $\lambda$, $\gamma$ and others will be found to flow
upon renormalisation, with the initial values $\lambda(0)=1$, $\gamma(0)=\gamma_0$, and
$K$ being the running momentum cutoff, which starts at $K=K_0$.
In a Fermi liquid $\lambda$ would
correspond to the inverse $Z$-factor, the quasiparticle weight.

{\it RG analysis.} 
{Perturbative treatment of disorder leads to
divergent contributions (with vanishing particle energy $E$)
to physical observables (conductivity, density of states, etc.).
These can be analysed using an RG approach, which}  
consists in integrating out the modes with the largest momenta
$\bk$: $K^\prime<|\bk|<K$. The action is reproduced with a new momentum cutoff $K^\prime$, renormalised gap $\Delta(l)$,
and the parameters $\lambda(l)$ and $\gamma(l)$ running according to
\begin{align}
	\partial_l\lambda &=C_d\frac{\gamma}{a^2}\lambda,
	\label{lambdaflow}
	\\
	\partial_l\gamma &=\varepsilon \,  \gamma + \frac{4 C_d}{a^2} \gamma^2,
	\label{gammaflow}
\end{align}
where $l=\ln(K/K^\prime)$,
$C_d=S_d/(2\pi)^d$, $S_d$ is the area of a unit
sphere in a $d$-dimensional space.

Eqs.~(\ref{lambdaflow})-(\ref{gammaflow}) are the one-loop perturbative RG equations controlled
by the dimensionless measure of disorder
$\gamma a^{-2}\ll1$ and, therefore, break down when this parameter flows to 
a value of order unity. The RG flow is terminated 
if the ultraviolet cutoff $K$ reaches $1/L$, $L$ being the characteristic size of the sample, or
the value $K_m$, at which
the energy scale $a K_m^\alpha/\lambda(K_m)$ is of the order of the energy $E$.

If $\varepsilon>0$, $\gamma$ flows towards larger values in accordance with Eq.~(\ref{gammaflow}). However, for
$\varepsilon<0$, $\gamma$ flows
to larger values if initially $\gamma>\gamma_c$, and flows to zero if $\gamma<\gamma_c$, where 
\begin{equation}
	\gamma_c = -\varepsilon (4 C_d)^{-1} a^2	
	\label{gammac}
\end{equation}
is the critical fixed point at which $\gamma$ does not flow. 
We expect, as is common in the study of critical phenomena, that such a critical point exists even if $\varepsilon$ is not 
small.

We note, that in addition to the random potential, considered here, there exist
other types of disorder, which we do not consider and which lead to an RG equation similar to Eq.~(\ref{gammaflow}),
but with a negative coefficient before the $\gamma^2$-term
(for example, 2D Dirac fermions with random-mass disorder).
In that case, the disorder strength flows towards smaller values above the critical dimensions ($\varepsilon<0$)
and has an attractive fixed point otherwise\cite{DotsenkoDotsenko,LudwigFisher}.

{
{\it Qualitative interpretation.}
Effectively the RG coarse-grains over the random disorder
potential on the scale of a wavelength $k^{-1}$, interpreting a complex of underlying impurities on
smaller scale as an effective impurity generating a renormalized random potential.
The structure of the linear (in $\gamma$) term in the beta-function in Eq.~(\ref{gammaflow}) and the
existence of the critical dimension
can be understood qualitatively by comparing 
the typical value $\gamma_0^\frac{1}{2}K_0^\frac{\varepsilon}{2}k^\frac{d}{2}$
of the average disorder potential in volume $k^{-d}$, with 
the kinetic energy $ak^\alpha$ for momentum $k$.
If $d<2\alpha$, the relative strength of disorder grows as $k\rightarrow0$.
In contrast, for $d>2\alpha$ the typical potential decreases at low momenta relative to 
the kinetic energy.
Carrying this coarse-graining procedure to higher (e.g., second) orders\cite{AleinerEfetov} in the disorder strength
(e.g., by replacing a pair of impurities, separated by $\lesssim k^{-1}$,
by an effective impurity) one arrives at the $\gamma^2$ and higher-order terms in the RG flow
equation for the disorder strength.
%
%
}

{\it Solutions.}
To analyse the low-energy behaviour of the conductivity, we solve 
Eqs.~(\ref{lambdaflow}) and (\ref{gammaflow}) with the result
\begin{align}
	&\gamma(K)K^\varepsilon=
	\frac{\gamma_0 K_0^\varepsilon}{1-\gamma_0/\gamma_c+(\gamma_0/\gamma_c)(K_0/K)^\varepsilon},
	\label{gammaanswer}
	\\
	&\lambda(K)=[\gamma(K) K^\varepsilon]^{1/4}(\gamma_0 K_0^\varepsilon)^{-1/4}.
	\label{lambdasolution}
\end{align}

At $K=K_{m}$, when
the RG stops, one arrives at an effective low-energy theory with a renormalised
action, which can be used further to evaluate physical
observables (conductivity, heat capacitance, magnetic susceptibility, etc.), e.g., in the usual Fermi-liquid approximation.

We now apply the above analysis of the renormalised field theory 
[Eqs.~(\ref{eq:gf})-(\ref{intL}), (\ref{gammaanswer}), (\ref{lambdasolution})]
 to the 
conductivity of a variety of systems.
For a finite doping in the conduction band, corresponding to the Fermi momentum $K_m$,
the Drude contribution\cite{AGD} to the conductivity is given by
\begin{equation}
	\sigma(K_m)=\frac{v^2(K_{m})}{2\pi\gamma(K_{m}) K_{m}^\varepsilon d}
	=\frac{\alpha^2a^2 K_{m}^{2\alpha-2}}{2\pi\gamma(K_{m})K_{m}^\varepsilon d},
	\label{DrudeDoping}
\end{equation}
where
$v(K_{m})$ is the velocity.
The Drude formula neglects weak-localisation effects
and accurately describes the conductivity only when they
are small and the disorder is weak, $\gamma(K) a^2\ll1$.

{\it Relevant disorder.} Let us consider the case of lower than critical dimensions,
$\varepsilon>0$. This is realised, for example, in
{conventional 2D and 3D semiconductors} with a quadratic dispersion ($\alpha=2$)
near the bottom
of the conduction band (the top of the valence band),
Fig.~\ref{PhaseDiagram}.
At $\varepsilon>0$ 
the disorder strength, Eq.~(\ref{gammaanswer}), grows upon
renormalisation and
diverges at a finite momentum cutoff
\begin{equation}
	K_{loc}=K_0(1-\gamma_c/\gamma_0)^{-1/\varepsilon}.
	\label{Kmobility}
\end{equation}
The singularity in the disorder strength in Eq.~(\ref{gammaanswer}), 
\begin{equation}
	\gamma(K) K^\varepsilon\propto(K-K_{loc})^{-1},
	\label{GammaSingularity}
\end{equation}
signals of the mobility threshold at
the momentum (\ref{Kmobility}). 
Strictly speaking, our calculation is not a proof of the localisation of the states with momenta $k<K_{loc}$,
because the perturbative RG has to be stopped when the disorder strength becomes too large, $\gamma/a^2\sim 1$.
At momenta $k\lesssim K_{loc}$ transport and localisation have to be studied by means of other
techniques, such as nonlinear sigma-model\cite{Efetov:book}, derived from our renormalised effective action.

For sufficiently large temperature $T$, the RG flow is terminated at energies $E\sim T$,
while the disorder is still weak, $\gamma a^{-2}\ll1$. The respective cutoff momentum $K$ is determined by
the condition
$a K^\alpha\sim\lambda(K)T$. In this case the conductivity remains finite and sufficiently large,
$\sigma[K(T)]>\sigma(K^*)$, where $K^*$ is the value of the momentum at which the perturbative RG breaks down,
$\gamma(K^*)a^{-2}\sim 1$.

For small finite doping in the conduction band and a large forbidden band $\Delta\gg T$,
the charge carriers are described by Boltzmann
statistics with the distribution function $n_F(E)\propto T^{-d/\alpha}e^{-E/T}$.
Using Eqs.~(\ref{Kubo}), (\ref{lambdasolution}) and (\ref{DrudeDoping}),
we estimate 
\begin{equation}
	\sigma(T)\propto T^{4-d/\alpha}.
	\label{sigmaT4}
\end{equation}

At zero doping the conductivity is exponentially small, $\sigma\propto e^{-\Delta/T}$,
as the charge carriers
have to get thermally excited to the conduction band 
in order to contribute to transport.

{\it Critical points of the disorder strength.} When the dimensionality of space is above its critical value,
$\varepsilon<0$, the flow of $\gamma$ has a critical point $\gamma_c$.
Near the critical point the dependency of the conductivity 
on $\gamma_c-\gamma_0$, $T$, and 
the chemical potential
can be understood from the standard scaling arguments\cite{CL}.

The characteristic wavelength $\xi$ of the charge carriers, which dominate the conductivity at $T=0$, scales with
small $\delta \gamma = \gamma_c-\gamma_0$ as $\xi\propto|\delta\gamma|^{-\nu}$.
It can be shown that the scaling of the conductivity at $T=0$ is given by the dimensional analysis,
$\sigma\propto\xi^{2-d}$. This leads to the scaling form of the conductivity
\begin{equation}
	 \sigma^{E}(\delta \gamma) \sim
	 |\delta \gamma|^{\nu(d-2)} g\left[{(E-\Delta)}{|\delta \gamma|^{-z \nu}}\right],
	 \label{eq:sigmascaling}
\end{equation}
where $z$ is the dynamic critical exponent\cite{CL}, $g$ is a scaling function (which, in general, depends
on the sign of $\delta\gamma$), and $E>\Delta$ (e.g., due to doping).
The conductivity at zero doping and finite
temperature can be obtained by averaging $\sigma^E(\delta \gamma)$ wrt $E$
with the distribution function $n_F^\prime(E)$, yielding
\begin{equation}
	 \sigma(\delta \gamma, T) \sim
	 T^\zeta|\delta \gamma|^{\nu(d-2)}  \tilde g\left[{T}{|\delta \gamma|^{-z \nu}}\right],
	 \label{eq:sigmascalingT}
\end{equation}
where $\tilde g$ is another scaling function,
$\zeta=0$ for gapless semiconductors ($\Delta\ll T$) and $\zeta=-d/\alpha$ for gapped
semiconductors ($\Delta\gg T$).

In particular, at zero temperature in a gapless weakly doped material
$\sigma \propto |\gamma_c- \gamma_0|^{\nu(d-2)}$.
At the critical point $\gamma=\gamma_c$, 
$\sigma(T) \propto T^{({d-2})/{z}}$ 
and $\sigma(T) \propto T^{({d-2})/{z}-d/\alpha}$ for gapless and gapped semiconductors
respectively.


{\it Dirac-type quasiparticles.}
The case of higher than critical dimensions, $d>2\alpha$, may be realised in WSMs (cf. Fig.~\ref{PhaseDiagram}),
3D materials with Dirac quasiparticle
spectrum\cite{Burkov:WeylProp,Wan:WeylProp,Liu:Na3Bi,Hasan:Cd3As2,Cava:Cd3As2,Yazdani:Cd3As2,Liu:Cd3As2} (with $\alpha=1$)
$
	\xi_\bk^{Weyl}=v\hbsigma\cdot\bk,
$
where the ``pseudospin'' $\hbsigma$ is the vector of Pauli matrices.
The quasiparticle Hamiltonian may also have a non-trivial sublattice and valley structure,
which has to be properly taken into account.
For simplicity, we focus on long-scale disorder confining the analysis near a single Weyl point.


In 2D our $\varepsilon = 0$ RG flow, Eqs.~(\ref{lambdaflow})-(\ref{gammaflow}),
reduces to that of Ref.~\cite{AleinerEfetov} for the strongest long-wavelength  disorder
in graphene (neglecting the other four types of disorder), cf. Fig.~\ref{PhaseDiagram}.

%

An equation similar to Eq.~(\ref{gammaflow}) for Weyl fermions has been also derived in Refs.~\onlinecite{Fradkin1}
{and~\onlinecite{GoswamiChakravarti}}.
In contradiction to our findings, {in Ref.~\onlinecite{Fradkin1}}
the conductivity has been found to vanish at weak disorder. We attribute this
to the shortcoming of the large N (number of valleys) approximation, equivalent to the self-consistent Born approximation (SCBA) (also used recently in Ref.~\onlinecite{Ominato:WeylDrude}), which does not account properly for the renormalisation effects found here (for the criticism of the SCBA see Refs.~\onlinecite{AleinerEfetov}
and \onlinecite{OstrovskyGornyMirlin}).

In order to generalise our RG approach to a WSM, 
we analyse the quasiparticle Hamiltonian of the form
\begin{equation}
	\xi_\bk=v k^{\frac{1}{2}+\frac{\varepsilon}{2}}\hbsigma\cdot\bk.
	\label{HamGen}
\end{equation}
At $\varepsilon=0$ it corresponds to $\alpha=3/2$, in which case $d=3$ is the critical dimension.
At $\varepsilon=-1$ the spectrum (\ref{HamGen}) turns into that of a WSM.
So, in order to address conduction in a 3D Weyl semimetal, we carry out the RG analysis
for the dispersion (\ref{HamGen}) at small $\varepsilon$ and then in the spirit of $\varepsilon$-expansion set
$\varepsilon=-1$. 

\begin{figure}[htbp]
	\centering
	\includegraphics[width=0.4\textwidth]{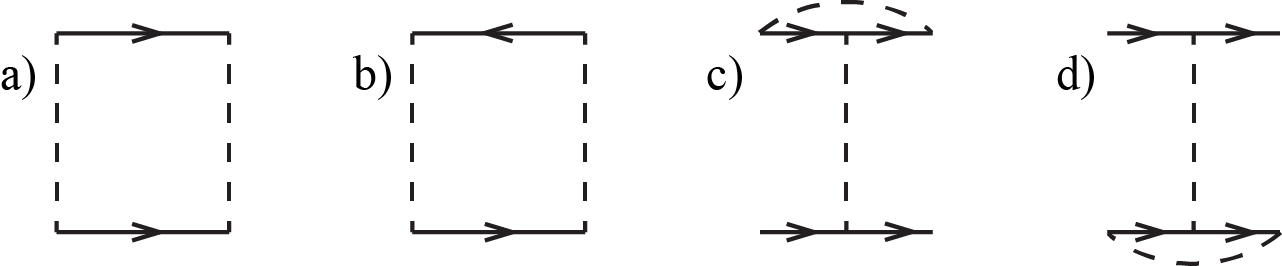}
	\caption{
	\label{GammaDiagr} Diagrams for the renormalisation of the disorder strength.
	}
\end{figure}

Repeating the above calculation with the spectrum (\ref{HamGen}), we arrive at the same
RG equations (\ref{lambdaflow})-(\ref{gammaflow}) with the
factor $4$ (corresponding to the four equally-weighted diagrams in Fig.~\ref{GammaDiagr}) in Eq.~(\ref{gammaflow})
replaced by $2$ (corresponding to the diagrams a) and b) cancelling each other for Dirac-type quasiparticles\cite{OstrovskyGornyMirlin}).
This leads to the doubling of the critical disorder strength $\gamma_c$, Eq.~(\ref{gammac}), and the exponent,
$1/4\rightarrow1/2$, in Eq.~(\ref{lambdasolution}).
In 3D $\gamma_c=\pi^2v^2$.

Thus, the presence of the pseudospin does not modify qualitatively the structure of the RG equations and their
solutions, but only changes coefficients of order unity. 

{\it Conductivity of Weyl semimetals.}
{
The RG equations (\ref{lambdaflow}), (\ref{gammaflow}) yield the values of the critical exponents\cite{GoswamiChakravarti}}
\be \nu = - \varepsilon^{-1}, \ z = 3/2.
\label{WeylExp}
\ee
For $\varepsilon=-1$, $\nu=1$.  

The Drude conductivity of a WSM \cite{Ominato:WeylDrude,RyuBiswas,Burkov:WeylProp,CompleteRubbish}
\footnote{See Supplemental Material for details.}
\begin{equation}
	\sigma=\frac{v^2}{2\pi\gamma(K) K^{-1}}
	\label{WeylDrude}
\end{equation}
with renormalised disorder strength $\gamma(K) K^{-1}$
is again suppressed for $\gamma>\gamma_c$ at low energies,
and remains large for $\gamma<\gamma_c$ [26].


The elastic scattering time at momentum $K$ 
\begin{equation}
	\tau(K)=\frac{1}{\pi\nu(K)\gamma(K)K^{-1}}=\frac{2\pi v}{K^2\left[\gamma(K)K^{-1}\right]}
	\label{Kdiverge}
\end{equation}
diverges $\propto K^{-2}$ at small momenta $K\rightarrow0$, as the
disorder strength $\gamma(K)K^{-1}$ saturates at a constant value for $\gamma_0<\gamma_c$, (\ref{gammaanswer}).
The divergent scattering time $\tau(K)$ ensures a finite conductivity $\sigma\sim v^2\nu(K)\tau(K)$
at low energies, despite the vanishing
density of states $\nu(K)$.
{We note, that at very small momenta $K<K_{rare}$ the conductivity may be dominated
by non-perturbative effects from exponentially rare spatial regions\cite{Nandkishore:rareregion}.}

Eq.~(\ref{Kdiverge}) implies that the parameter $Kv\tau(K)$ remains large
under the RG for $\gamma<\gamma_c$ if it was so in the bare system. 
Then one may neglect diagrams with crossed impurity lines\cite{AGD}
in the system with renormalised parameters, and the weak-localisation corrections 
to the Drude conductivity of a 3D material are small. 

Thus, at weak disorder, the Drude conductivity in terms of the renormalised parameters,
Eq.~(\ref{WeylDrude}), accurately describes the full
conductivity of WSM. Moreover, 
since $\gamma(K)K^{-1}$ saturates at a constant for low energies, 
Eq.~(\ref{WeylDrude}) also describes
 the conductivity
at zero doping
and finite temperature, {$T\gg\tau^{-1}, vK_{rare}$}.


To compute the conductivity at $\gamma_0<\gamma_c$, we use Eq.~(\ref{WeylDrude}) with
the renormalised disorder strength $\gamma(K_m)K_m^{-1}$, given by Eqs.~(\ref{gammaanswer}), (\ref{lambdasolution})
[with the aforementioned $1/4 \rightarrow 1/2$ replacement],
and the flow terminated by the cutoff $K_m$, set by $vK_m\sim\lambda(K_m) T$.

We find 
\begin{equation}
	\sigma_0(T)=\frac{v^2K_0}{2\pi\gamma_0}
	\left(
	1-\frac{\gamma_0}{\gamma_c}+\frac{\gamma_0}{\gamma_c}\frac{T}{W}
	\right)
	\label{CondWeylT}
\end{equation}
for $T\ll W$, where $W$ is a constant of the order of $K_0v(1-\gamma_0/\gamma_c)^{1/2}$.
Eq.~(\ref{CondWeylT}) is consistent with the scaling theory, Eq.~(\ref{eq:sigmascalingT}),
with the scaling exponents (\ref{WeylExp}).

For $\gamma_0>\gamma_c$, the system flows to the strong disorder regime. 
As discussed above, 
the RG may be terminated by sufficiently high 
temperature, also ensuring that the weak-localisation corrections remain small.
Similarly to Eq.~(\ref{sigmaT4}), we find $\sigma(T)\sim  \gamma_0^{-1}K_0 (T/K_{loc})^2\propto T^2\delta\gamma^{-2}$.
Close to the critical point 
Eq.~(\ref{eq:sigmascalingT}) 
yields
$\sigma(T)\propto T^{2/3}$.

At low temperatures, zero doping, and $\gamma>\gamma_c$ the RG breaks down when
the disorder becomes strong, $\gamma\sim v^2$, 
equivalent to the criterion when weak-localisation corrections become important.
At the breakdown point the system has no small parameters and is characterised
by momentum scale $K^*$.
Although the conductivity is strongly suppressed, {\it single}-node
Weyl fermions are topologically protected from localisation\cite{Wan:WeylProp, RyuLudwig:classification}.

Assuming a finite conductivity $\sigma$ in this regime, 
we estimate, using Eq.~(\ref{gammaanswer}),
\begin{equation}
	\sigma^*\sim K_0\left(\kappa v^{-2}-\gamma_c^{-1}\right)^{-1}\left(\gamma_0^{-1}-\gamma_c^{-1}\right),
	\label{SigmaBD}
\end{equation}
where $\kappa$ is a constant of order unity.
The linear dependency of the conductivity on the disorder strength, $\sigma^*\propto\gamma_0-\gamma_c$ near
the critical point is consistent with the predictions of the scaling theory, Eqs.~(\ref{eq:sigmascalingT})
and (\ref{WeylExp}).

{
{\it Experimental implications.} 
Recently, Dirac (Weyl) quasiparticle dispersion has
been reported\cite{Liu:Na3Bi,Hasan:Cd3As2,Cava:Cd3As2,Yazdani:Cd3As2,Liu:Cd3As2} in $Cd_3As_2$
and $Na_3Bi$, which possibly present a platform for observing the conductivity dependency $\sigma(\gamma,T)$,
Eqs.~(\ref{CondWeylT})-(\ref{SigmaBD}) and Fig.~\ref{WeylCondSurface}, which we predict.
Our results apply to
Weyl materials with short-range-correlated disorder (e.g., neutral impurities or vacancies)
slightly doped away from the Weyl point or at finite temperatures. 
}



{\it Acknowledgements.} 
{We are grateful to L.~Balents, E.~Bergholtz, P.~Brouwer,
R.~Nandkishore, and B.~Sbierski
for useful discussions and remarks on the manuscript.}
Our work has been supported by the Alexander von Humboldt
Foundation through the Feodor Lynen Research Fellowship (SVS) and by the NSF grants
DMR-1001240 (LR and SVS),
DMR-1205303 (VG and SVS), PHY-1211914 (VG and SVS), and PHY-1125844 (SVS).


{\it Note added.} While this work was under review in Physical Review Letters, another paper, Ref.~\cite{Brouwer:WSMcond}, was submitted and published, numerically addressing conductivity of WSM at zero doping. 
We present a detailed comparison of our results for a WSM with those of Ref.~\cite{Brouwer:WSMcond} in the Supplemental
Material.



\newpage

\renewcommand{\theequation}{S\arabic{equation}}
\renewcommand{\thefigure}{S\arabic{figure}}
\renewcommand{\thetable}{S\arabic{table}}
\renewcommand{\thetable}{S\arabic{table}}
\renewcommand{\bibnumfmt}[1]{[S#1]}
\renewcommand{\citenumfont}[1]{S#1}

\setcounter{equation}{0}
\setcounter{figure}{0}
\setcounter{enumiv}{0}

\onecolumngrid
\begin{center}
\textbf{\large Supplemental Material for \\
``Critical transport in weakly disordered semiconductors and semimetals''}
\end{center}
\vspace{2ex}
\twocolumngrid

\section{Renormalisation of the band gap}

From Eqs.~(3)-(5) we find that the quantity $\lambda\Delta$ flows upon renormalisation as 
\begin{equation}
	\partial_l(\lambda\Delta)=-C_dK^\alpha\frac{\gamma}{a}.
	\label{DeltaLambdaFlow}
\end{equation}
This can be absorbed into the redefinition of the chemical potential
or the band gap $\Delta$, which then flows as
\begin{equation}
	\partial_l\Delta=-C_dK^\alpha\frac{\gamma}{a\lambda}-C_d\Delta\frac{\gamma}{a^2}.
	\label{DeltaFlow}
\end{equation}

The renormalisation of the band gap by the elastic scattering processes
in the conduction band is similar to the renormalisation of the critical temperature in
the $\phi^4$-theory.

As Eqs.~(\ref{DeltaLambdaFlow}) and (\ref{DeltaFlow}) explicitly contain the ultraviolet
momentum cutoff $K$ in the right-hand-side part, the exact value of the renormalised
band gap $\Delta$ depends on the details of the cutoff procedure.

\section{Drude conductivity of Weyl semimetal}

\subsubsection*{Weak disorder, $\gamma\ll\gamma_c$}

Let us first evaluate the Drude contribution to the conductivity of Weyl semimetal for a finite chemical
potential $\mu>0$ (measured from the bottom of the conduction band) and in the limit of weak
disorder, $\gamma_0\ll\gamma_c$. As we have demonstrated, in this case the disorder strength $\gamma(K)K^{-1}$ 
does not experience renormalisations, and the interference phenomena far from the Fermi surface
can be neglected. 
Then the conductivity is determined by the scattering between the momentum states with energies close to the
chemical potential $\mu$.

Because of the small disorder strength, the elastic scattering rate can be
evaluated in the Born approximation;
\begin{equation}
	\tau^{-1}=2i\gamma(K)K^\varepsilon\int \langle\hat G^R(\mu,\bk)\rangle_{dis}
	\frac{d\bk}{(2\pi)^3}=\pi\rho_F\gamma(K)K^{-1},
	\label{tauBorn}
\end{equation}
where $\rho_F(\mu)=\frac{\mu^2}{2\pi^2v^3}$ is the density of states in Weyl semimetal,
and
\begin{equation}
	\langle\hat G^{R}(\mu,\bk)\rangle_{dis}=\frac{\mu{\hat 1}+v\hbsigma\bk}{(\mu+\frac{i}{2\tau})^2-v^2k^2},
	\label{GRaveraged}
\end{equation}
is the disorder-averaged retarded Green's function, a $2\times2$ matrix in the pseudospin space,
with $\hat 1$ being the unity matrix.

\begin{figure}[htbp]
	\centering
	\includegraphics[width=0.3\textwidth]{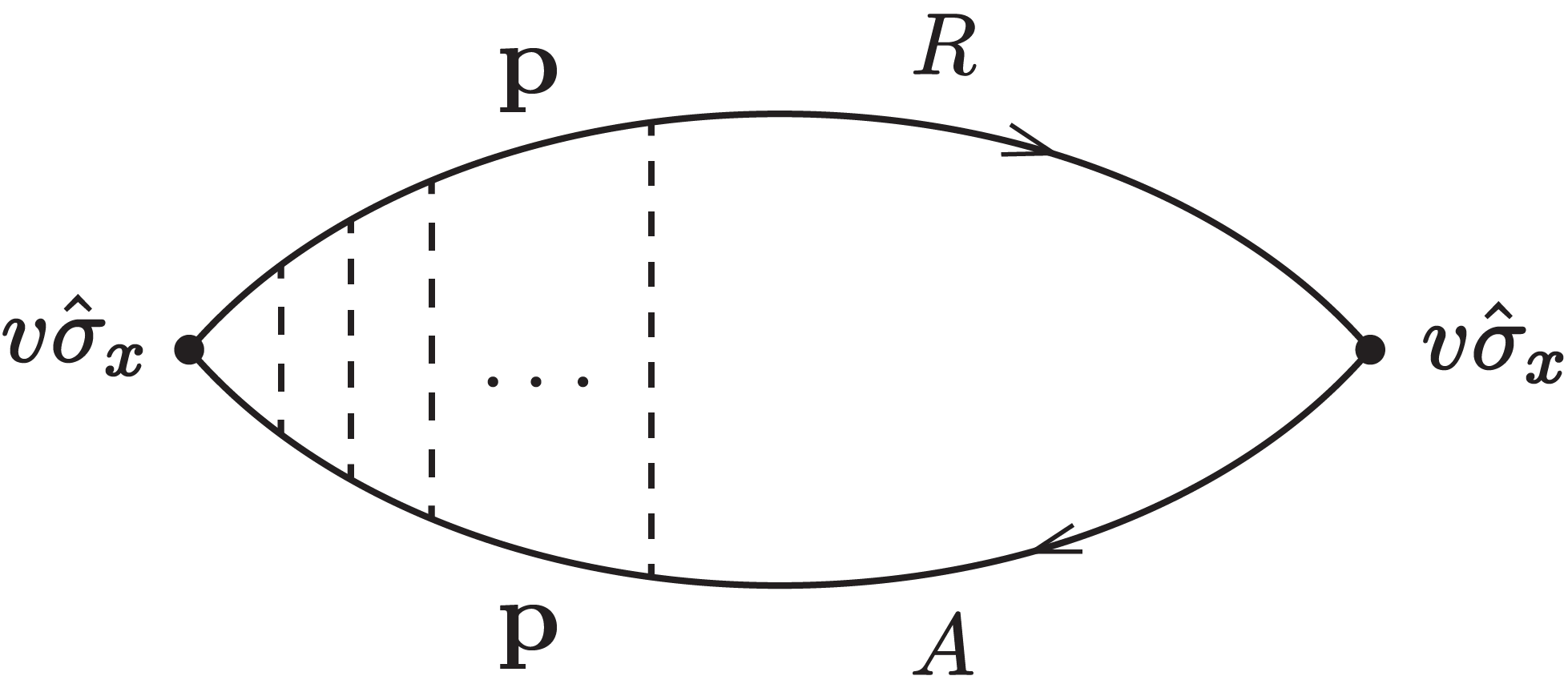}
	\caption{\label{Fish}
	Diagram for the Drude conductivity in Weyl semimetal.
	}
\end{figure}

The conductivity is given by the Kubo formula
\begin{equation}
	\sigma_{xx}=\frac{v^2}{2\pi}
	\int{\mathrm Tr\left<\hat\sigma_x \hat G^A(\mu,\br,\br^\prime) \hat\sigma_x \hat G^R(\mu,\br^\prime,\br)\right>_{dis}}
	d\br^\prime,
	\label{KuboWeyl}
\end{equation}
where $\hat G^R(\mu,\br^\prime,\br)$ and $\hat G^A(\mu,\br,\br^\prime)$
are the retarded and advanced Green's functions, 
and $v\hat\sigma_x$ is the operator of velocity along the $x$ axis.
In the limit of weak disorder under consideration the conductivity is dominated by
the Drude contribution, shown diagrammatically in Fig.~\ref{Fish}, and the weak-localisation
corrections are negligible.

Each step of the diffusion ladder (diffuson), which renormalises the velocity vertex $v\hsigma_x$
in Fig.~\ref{Fish}, equals
\begin{align}
	\hat\Pi(\mu)=&(\pi\rho_F\tau)^{-1}\int \langle\hat G^R(\mu,\bp)\rangle_{dis}
	\otimes
	\langle\hat G^A(\mu,\bp)\rangle_{dis}\frac{d\bp}{(2\pi)^3}
	\nonumber\\
	&=\frac{1}{2}\left(1+\frac{1}{3}\Sigma_{i=1}^3\hsigma_i\otimes\hsigma_i\right),
	\label{Step}
\end{align}
where $\otimes$ is a product of the operators which act in the advanced and retarted spaces, i.e.
on the upper and the lower lines in Fig.~\ref{Fish},
and the prefactor $(\pi\rho_F\tau)^{-1}$ is the value of the impurity line.

Using Eq.~(\ref{Step}), we sum up the diffusion
ladder in Fig.~\ref{Fish} and obtain the renormalised velocity vertex:
\begin{align}
	\tilde v\hsigma_x=v\hsigma_x+\frac{v}{2}\left(\hsigma_x+\frac{1}{3}\sum_{i=1}^3\hsigma_i\hsigma_x\hsigma_i\right)+\ldots
	=\frac{3}{2}v\hsigma_x.
\end{align}

The conductivity is then given by
\begin{equation}
	\sigma=\frac{v^2\rho_F\tau}{2}=\frac{v^2}{2\pi\gamma(K)K^{-1}}.
	\label{WeylDrudeSuppl}
\end{equation}
The conductivity~(\ref{WeylDrudeSuppl}) is independent of the chemical potential and momentum $K$,
as the disorder strength $\gamma(K)K^{-1}=\gamma_0 K_0^{-1}$ does not depend on $K$ in the limit of
weak disorder.

The Drude contribution to the conductivity at weak disorder, Eq.~(\ref{WeylDrudeSuppl}),
has been calculated recently in Ref.~\onlinecite{SOminato:WeylDrude} using the kinetic-equation approach and,
diagrammatically, in Ref.~\onlinecite{SRyuBiswas}.
Similar formulas have been obtained also in Refs.~\onlinecite{SBurkov:WeylProp} and \onlinecite{SCompleteRubbish};
however, the renormalisation of the velocity vertex by a diffuson, Fig.~\ref{Fish}, has not been taken into account,
leading to a mistake of order unity in the expression for conductivity.

\subsection*{Renormalised disorder}

For $\gamma_0$ of the order of or larger than $\gamma_c$ the system is subject to strong
renormalisation. The RG procedure removes the higher momenta from the system, resulting in a renormalised
action of the quasiparticles near the Fermi surface.

If $\gamma_0<\gamma_c$, the system flows towards weaker disorder, and the conductivity can be
then evaluated as above, in a controlled weak-disorder perturbation theory,
using the renormalised chemical potential $\mu^\prime=\mu\lambda(K)$, and the disorder strength 
$\gamma(K)K^{-1}$, where $vK=\mu^\prime$. As the conductivity is independent of the chemical
potential $\mu^\prime$, it is given by Eq.~(\ref{WeylDrudeSuppl}) with the renormalised disorder
strength $\gamma(K)K^{-1}$.

In the case $\gamma_0>\gamma_c$ the system flows towards strong disorder. Eq.~(\ref{WeylDrudeSuppl})
can still be applied if the RG flow is terminated by a sufficiently large Fermi energy, such that 
the effective disorder remains weak, $\gamma(K)\ll v^2$.


\section{Self-consistent Born approximation}

In this Section we analyse transport in a high-dimensional semiconductor
by means of the self-consistent Born approximation (SCBA) and discuss the difference
between the SCBA results and those obtained from the RG analysis 
controlled by an $\varepsilon=2\alpha-d$-expansion.

The imaginary part $\Sigma(E)$
of the self-energy of, e.g., the disorder-averaged retarded Green's function can be obtained 
for short-range disorder within the
SCBA using the self-consistency equation
\begin{equation}
	\Sigma(E)=\gamma_0 K_0^\varepsilon	\int\frac{d\bp}{(2\pi)^d}
	\frac{\Sigma(E)}{(E-a p^\alpha)^2+\Sigma^2(E)}.
	\label{SCE}
\end{equation}

\begin{figure}[ht]
	\centering
	\includegraphics[width=0.45\textwidth]{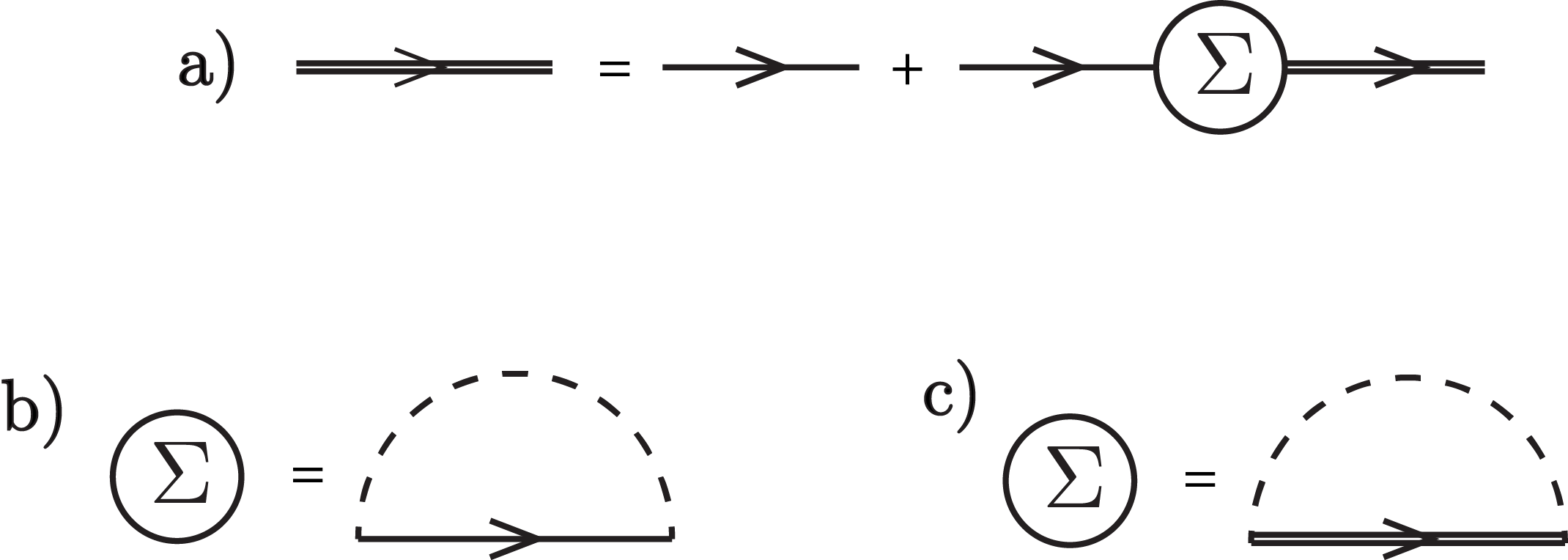}
	\caption{\label{Fig:SCBAdiagr}
	Diagrams for a disorder-averaged Green's function. Double and single solid lines
	correspond to disorder-averaged and bare Green's functions respectively. 
	a) Dyson equation. b) Self-energy part in the Born approximation.
	c) Self-energy part in the self-consistent Born approximation.
	}
\end{figure}

Let us first address the behaviour of the self-energy part of the zero-energy ($E=0$)
Green's function.

Because in high dimensions ($\varepsilon=2\alpha-d<0$)
\begin{equation}\int\frac{d\bp}{(2\pi)^d}
	\frac{1}{a^2p^{2\alpha}+\Sigma^2(0)}
	\leq\int\frac{d\bp}{(2\pi)^d}\frac{1}{a^2p^{2\alpha}}=\frac{C_dK_0^{|\varepsilon|}}{|\varepsilon| a^2},
\end{equation}
Eq.~(\ref{SCE}) has a non-zero solution $\Sigma(0)\neq0$ only for sufficiently strong
disorder $\gamma_0>\gamma_c^{SCBA}$,
with
\begin{equation}
	\gamma_c^{SCBA}=\frac{|\varepsilon| a^2}{C_d}\equiv 4\gamma_c.
\end{equation}

Thus, the SCBA analysis suggests that states at the bottom of the band undergo a phase transition
at $\gamma_0=\gamma_c^{SCBA}$; for supercritical disorder the elastic scattering time
$\tau(0)=[2\Sigma(0)]^{-1}$ is finite, while for disorder below critical the system is
effectively ballistic, $\tau(0)=[2\Sigma(0)]^{-1}=\infty$. 

Although the SCBA correctly predicts the existence of the disorder-driven phase transition,
the critical value of the disorder strength $\gamma_c^{SCBA}=4\gamma_c$, obtained from the SCBA,
is different by a factor of $4$ from the value $\gamma_c$, obtained from the controlled-in-$\varepsilon$
RG analysis.

\begin{figure}[ht]
	\centering
	\includegraphics[width=0.45\textwidth]{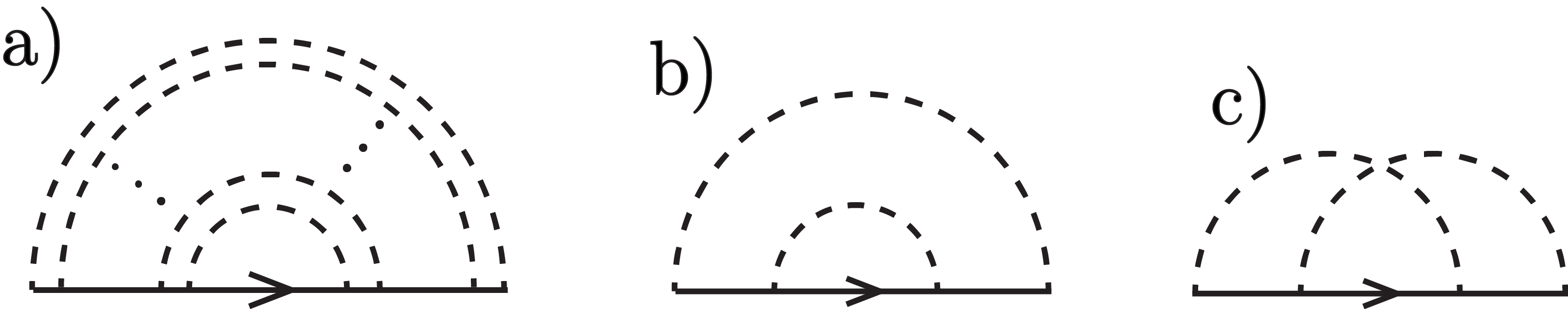}
	\caption{
	\label{Fig:SelfEnergyDiagr}
	Diagrams contributing to the self-energy part.
	a) ``Rainbow'' diagrams, contributing in the SCBA.
	b) SCBA contribution in the fourth order in the random potential.
	c) A contribution of the same order as b) neglected in the SCBA. 	
	}
\end{figure}

Indeed, the SCBA self-energy part is given by the sum of ``rainbow'' diagrams, Fig.~\ref{Fig:SelfEnergyDiagr},
and thus does not take into account the other diagrams that contribute to the quasiparticle self-energy
part. For instance in the fourth order in the disorder potential the SCBA takes into account the diagram
in Fig.~\ref{Fig:SelfEnergyDiagr}b and disregards that in Fig.~\ref{Fig:SelfEnergyDiagr}c, which 
for small momenta and energy $E$ has the same order of magnitude of the imaginary part
\begin{equation}
	\Sigma^{3b}\sim\Sigma^{3c}\sim\frac{\gamma_0^2\rho_{clean}(E)}{|\varepsilon|a^2K_0^{|\varepsilon|}},
\end{equation}
where we have introduced the density of states in a clean semiconductor
\begin{equation}
	\rho_{clean}(E)=\frac{C_d E^\frac{d-\alpha}{\alpha}}{\alpha a^\frac{d}{\alpha}}.
	\label{RhoClean}
\end{equation}

We emphasise that, although diagram \ref{Fig:SelfEnergyDiagr}c contains crossing
impurity lines, in a higher-dimensional system under consideration 
it is not suppressed by the parameter $\left[E\tau(E)\right]^{-1}\ll1$, unlike the case of
a conventional metal\cite{SAGD}. Such suppression in low-dimensional metals and semiconductors
occurs if only momenta close to the Fermi surface are important\cite{SAGD}, while in
diagrams \ref{Fig:SelfEnergyDiagr}b and \ref{Fig:SelfEnergyDiagr}c the momentum
integration is carried out over all momenta in the band. 

Thus, the SCBA presents an uncontrolled approximation for studying transport in disordered
systems, except for sufficiently small disorder strengths, when it coincides with the usual
(non-self-consistent) Born approximation, that takes into account only the leading-order
contribution to the self-energy part. Detailed criticism of the applicability of the SCBA
to conduction in graphene is presented in Refs.~\cite{SAleinerEfetov} and \cite{SOstrovskyGornyMirlin}.
Also, SCBA in Weyl semimetal has been recently discussed in Ref.~\cite{SBrouwer:WSMcond}.

\subsection{Conductivity of a semiconductor in the SCBA}

{\it Scattering rate for subcritical disorder.}
For subcritical disorder, $\gamma_0<\gamma_c^{SCBA}$, the scattering rate $\Sigma(E)$
vanishes at the bottom of the band, $E=0$, but has a finite value for any $E>0$.

Assuming $E\gg\Sigma(E)$, the integral in the right-hand-side of Eq.~(\ref{SCE}) can be evaluated as a sum
of the contribution from large momenta $p \sim K_0$ and the contribution near the Green's function
pole $E\approx ap^\alpha$, yielding
\begin{equation}
	\Sigma(E)=\frac{\pi \rho_{clean}(E)\gamma_0 K_0^\varepsilon}{1-\gamma_0/\gamma_c^{SCBA}}.
	\label{SigmaSmall}
\end{equation}

Thus, for subcritical disorder ($\gamma_0<\gamma_c^{SCBA}$) the SCBA scattering rate $\Gamma$ decreases with energy
$\propto E^\frac{d-\alpha}{\alpha}$, faster than the energy $E$ in the case of higher dimensions under consideration
(where $\Gamma/E\propto E^{-|\varepsilon|}$, in agreement with the RG analysis of the weak disorder relevance),
and diverges as the disorder strength $\gamma_0$ approaches the critical strength $\gamma_c^{SCBA}$.

{\it Scattering rate for supercritical disorder.}
For stronger disorder, $\gamma_0>\gamma_c^{SCBA}$, the scattering rate $\Sigma(E)$ is finite already
at the bottom of the band.

For small $\varepsilon\ll1$ and $E=0$
the integral in the right-hand-side part of Eq.~(\ref{SCE}) can be evaluated
as
\begin{align}
	\int\frac{d\bp}{(2\pi)^d}
	\frac{1}{a^2p^{2\alpha}+\Sigma^2(0)}
	\nonumber \\
	=
	\int\frac{d\bp}{(2\pi)^d}\frac{1}{a^2p^{2\alpha}}
	-\int\frac{d\bp}{(2\pi)^d}\frac{\Sigma^2(0)}{a^2p^{2\alpha}\left[a^2p^{2\alpha}+\Sigma^2(0)\right]}
	\nonumber \\
	\approx\frac{C_dK_0^{|\varepsilon|}}{|\varepsilon| a^2}
	-\frac{C_dK_0^{|\varepsilon|}}{|\varepsilon| a^2}\left[\frac{\Sigma(0)}{aK_0^\alpha}\right]^\frac{|\varepsilon|}{\alpha}.
	\label{StrongSigmaCalc}
\end{align}
From Eqs.~(\ref{SCE}) and (\ref{StrongSigmaCalc}) we find that the scattering rate near the bottom of the band
for supercritical disorder is given by
\begin{equation}
	\Sigma(0)=aK_0^\alpha\left(1-\frac{\gamma_c^{SCBA}}{\gamma_0}\right)^\frac{\alpha}{|\varepsilon|}.
	\label{SigmaLarge}
\end{equation}

{\it Conductivity.}
In the case of a conventional semiconductor there is no renormalisation of the velocity vertex by the diffuson,
and the conductivity for the Fermi energy $E$ in the conduction band is given by
\begin{align}
	\sigma(E)=\frac{v^2(E)}{d}\int\frac{d\bp}{(2\pi^d)}G^R(\bp,E)G^A(\bp,E)
	\nonumber\\ 
	=\frac{\pi v^2\rho_{\text{clean}}(E)}{\Sigma(E)d}.
	\label{KuboSCBA}
\end{align}

Using Eqs.~(\ref{KuboSCBA}), (\ref{SigmaSmall}) and (\ref{SigmaLarge}), we obtain the
dependency of the conductivity on the Fermi level $E$ and the deviation
$\delta\gamma=\gamma_0-\gamma_c^{SCBA}$
of the disorder strength from the critical value:
\begin{equation}
	\sigma(E,\delta\gamma)
	\propto
	\left\{
	\begin{array}{cc}
		|\delta\gamma|\: E^{2-\frac{2}{\alpha}}, & \delta\gamma<0,
		\\
		E^\frac{d+\alpha-2}{\alpha} \delta\gamma^\frac{\alpha}{|\varepsilon|}, & \delta\gamma>0.
	\end{array}
	\right.
\end{equation}

Thus, for subcritical disorder ($\delta\gamma<0$) the SCBA predicts the same dependency of conductivity
on energy $E$ and disorder strength $\delta\gamma$
as the rigorous (for small $\varepsilon$) RG calculation [Eq.~(11) in the main
text].

However, for supercritical disorder ($\delta\gamma>0$) the SCBA is 
qualitatively incorrect: it predicts a {\it finite} conductivity
for a disordered semiconductor in the orthogonal symmetry class,
in which the conductivity is absent at strong disorder, $\sigma=0$, due to the localisation
of low-energy states\cite{SSyzranov:unconv}.

\subsection{Weyl semimetal and SCBA}

Transport in a Weyl semimetal can be studied by means of the SCBA similarly to the case of a semiconductor considered
above. The respective detailed SCBA analysis for WSM are presented in Ref.~\cite{SOminato:WeylDrude}.
Also, RG analysis in the large $N$ (number of valleys) approximation, equivalent to the SCBA, have been carried out
in Ref.~\onlinecite{SFradkin1}.

The self-consistency condition for the self-energy part $\Sigma(0)$ of a WSM is also given by Eq.~(\ref{SCE})
with $\alpha=1$, $a=v$, and $d=3$. It gives the critical disorder strength $\gamma_c=2\pi^2v^2$, which by a factor
of $2$ exceeds that obtained from the RG analysis with $\varepsilon=-1$ presented in this paper. For a small
finite chemical potential, the SCBA predicts a finite zero-temperature 
conductivity on both sides of the
transition, that vanishes at the critical point, which is qualitatively consistent with the result of the RG
analysis presented here.


\section{Conductance of a finite sample of Weyl semimetal}

Recently the conductivity of a finite sample of Weyl semimetal at zero temperature 
and chemical potential has been studied numerically in Ref.~\onlinecite{SBrouwer:WSMcond};
it has been concluded that the conductivity for strong disorder, $\gamma_0>\gamma_c$,
is finite and qualitatively consistent with the predictions of this paper,
while the conductivity 
for subcritical disorder, $\gamma_0<\gamma_c$,
vanishes.

At first glance, such vanishing weak-disorder conductivity
is inconsistent with our prediction of a finite conductivity
in the same regime. However, as explained in Ref.~\onlinecite{SBrouwer:WSMcond},
the conclusions of our paper apply to 
samples of sufficiently large sizes $L$ and finite temperatures or chemical potentials such that
$T, \mu\gg v/L$, while the results of Ref.~\onlinecite{SBrouwer:WSMcond} apply in the opposite
limit, and the qualitative difference between the results may come from the non-commutativity
of these limits.

\begin{figure}[ht]
	\centering
	\includegraphics[width=0.33\textwidth]{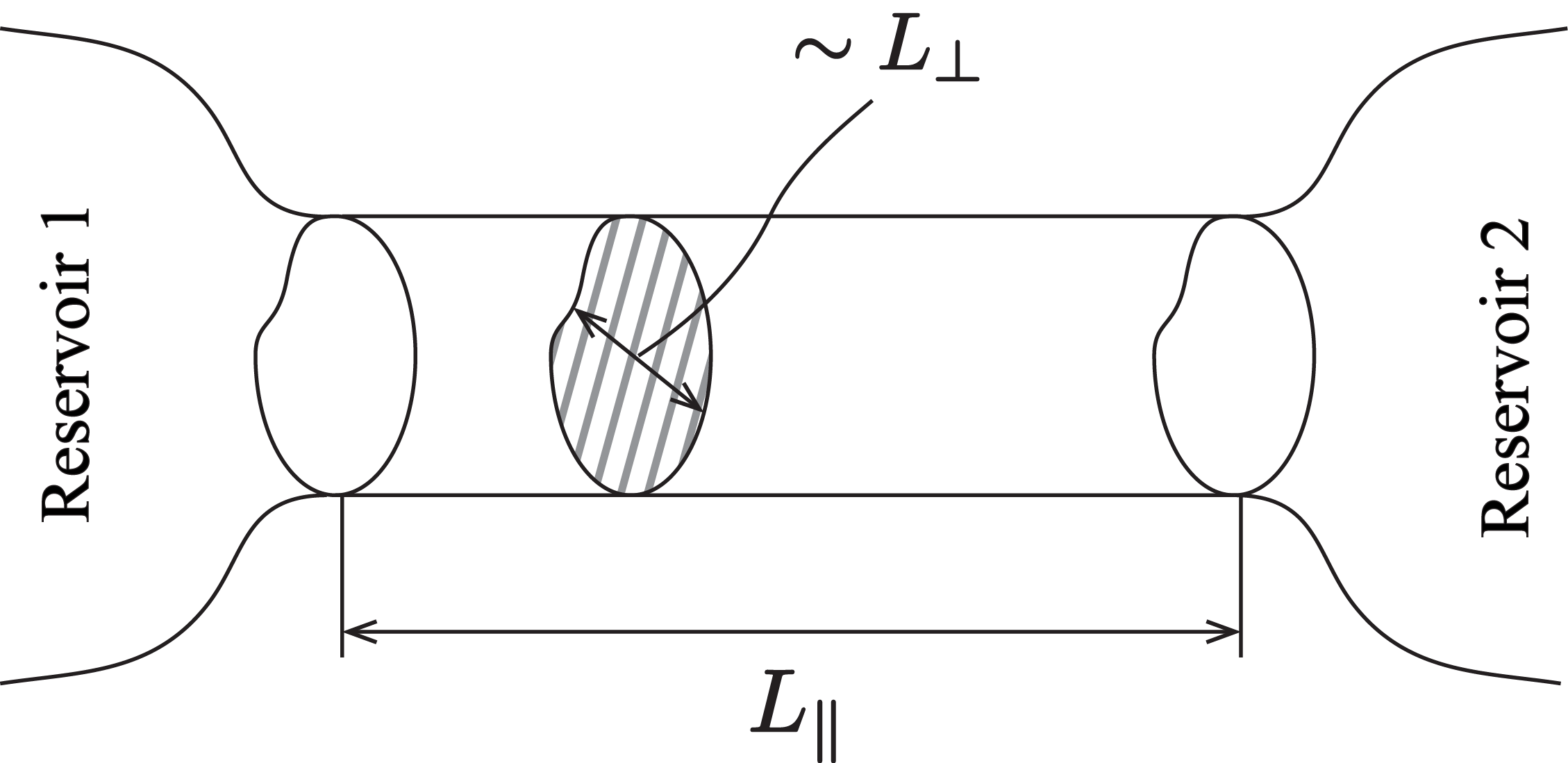}
	\caption{
	\label{Fig:setup}
	Setup for observing the conductance of a sample of length $L_\parallel$ and the characteristic
	width $L_\bot$, connected to two infinite reservoirs.
	}
\end{figure}

Indeed, in a disorder-free sample of finite width, Fig.~\ref{Fig:setup},
the transverse momentum of charge carriers is quantised, resulting in a finite number $N_\bot$ of
modes that contribute to the conductance. For instance, in a 3D sample with a finite cross-section $S_\bot$,
and sufficiently large Fermi momentum $k_F\gg L_\bot^{-1}\sim1/\sqrt{S_\bot}$ the number of modes (per one discrete degree
of freedom, e.g, spin and valley), that provide conduction at small temperatures $T\ll E_F$, is given by
\begin{equation}
	N_\bot= k_F^2S_\bot/\pi+{\cal O}(1).
\end{equation}

In an undoped Weyl semimetal, studied in Ref.~\onlinecite{SBrouwer:WSMcond}, $k_F=0$, and no more than one
or several transverse modes $N_\bot\sim1$ may contribute to conduction for {\it weak disorder} and temperatures
$T$ smaller than 
the characteristic energy scale $v/L_\bot$ of spatial quantisation. 
Therefore, the conductance of such system
is smaller or of the order of several conductance quanta, $G\lesssim1$.

Then the conductivity of such sample of length $L_\parallel$, defined according to
\begin{equation}
	\sigma=G\frac{L_\parallel}{S_\bot},
\end{equation}
vanishes, $\sigma\lesssim L_\parallel/L_\bot^2\rightarrow0$ in the limit $L_\parallel\propto L_\bot\rightarrow\infty$.

Thus, as emphasised in Ref.~\cite{SBrouwer:WSMcond}, the numerical zero-conductivity result,
obtained there for small disorder strengths, is likely attributable to considering finite sample size $L$ while setting the chemical
potential and temperature to zero.
However, in the limit $T,\mu \gg v/L$, the conductivity is finite and in the limit of weak disorder is described
by Eq.~\eqref{WeylDrudeSuppl},
thus resolving the apparent inconsistency between our predictions and Ref.~\onlinecite{SBrouwer:WSMcond}.

The above estimate for the conductivity is no longer valid in the case of {\it strong disorder}.
As we have shown recently in Ref.~\cite{SSyzranov:unconv},
the density of states at the Dirac point becomes finite at $\gamma>\gamma_c$ due to strong fluctuations of the
(renormalised) disorder potential $U^*\sim vK^*$, which thus leads to a large effective number
\begin{equation}
	N_\bot(\gamma>\gamma_c)\sim (K^*)^2S_\bot
\end{equation}
of conducting modes even in the absence of doping.
Although, due to backscattering, the contribution of each such mode to the
conductance is in general smaller than the conductance quantum, the conductivity may thus
become finite at strong disorder, as observed numerically in Ref.~\cite{SBrouwer:WSMcond}.


\end{document}